\documentclass[aps,prb,reprint,amsfonts,superscriptaddress]{revtex4-2}
\usepackage{amsmath}  
\usepackage{bm} 
\usepackage{graphicx}
\usepackage{gensymb}
\usepackage{color,soul}
\usepackage{makecell}
\usepackage{multirow}
\usepackage{soul}
\usepackage{adjustbox}
\graphicspath{ {./images/} }

\usepackage[dvipsnames]{xcolor}

\begin{document}

\title{Determination of the spin chirality using x-ray magnetic circular dichroism}

\author{Gerrit~\surname{van der Laan}}
\affiliation{Diamond Light Source, Harwell Science and Innovation Campus, Didcot, Oxfordshire OX11~0DE, United Kingdom}
% ORCID: 0000-0001-6852-2495

\date{\today}

\begin{abstract}
A 3-fold symmetric kagome lattice that has negative spin chirality can give a non-zero x-ray magnetic circular dichroism (XMCD) signal, despite that  the total spin moment amounts to zero. This is explained by a hitherto unnoticed rule for the rotational symmetry invariance of the XMCD signal. A necessary condition is the existence of an anisotropic XMCD signal for the local magnetic atom,
{\color{black} which can arise from a spin anisotropy either in the ground state or the final state}. 
The angular dependence of the XMCD as a function of the beam direction has an unusual behavior.
The maximum dichroism is not aligned along the spin direction, but  depends on the relative orientation of the spin with respect to the atomic orientation.
Therefore, different geometries can result in the same angular dependence, and the spin direction can only be determined if the atomic orientation is known.
The consequences for the x-ray magneto-optical sum rules are given. The integrated XMCD signals are proportional to the anisotropy in the orbital moment and the magnetic dipole term, where the isotropic spin moment drops out.
\end{abstract}

\maketitle

\section{Introduction}

X-ray magnetic circular dichroism (XMCD) has become a versatile technique to interrogate ferro- and ferri-magnetic magnetic materials \cite{vanderLaan2014}.
Particularly huge XMCD signals are observed in the soft x-ray region \cite{vanderLaan1991}, facilitating us to extract the expectation values of the element-specific spin and orbital moments \cite{Thole1992,Carra1993}.
On the other hand, common wisdom has it that there is no XMCD from antiferromagnetic (AFM) materials, where the spin moments for the particular element cancel each other \cite{Stohr2016}.

Recently, Yamasaki {\it{et al.}}\ \cite{Yamasaki2020} conjectured the  existence of XMCD in the coplanar 120$^\circ$ AFM kagome network of Mn$_3$Sn. These authors ascribed the origin of the XMCD to the magnetic dipole term, $\langle T_\zeta \rangle$. This term gives the anisotropy of the spin distribution  \cite{Carra1993} and  is contributing to the magnetocrystalline anisotropy energy \cite{vanderLaan1998jpcm}.  
Subsequently, Sasabe {\it et al.}~\cite{Sasabe2021} performed cluster calculations for the XMCD at the $L_{2.3}$  absorption edge of Fe$^{2+}$ $d^6$  in a kagome lattice, thereby theoretically confirming the presence of XMCD.
However, they also found XMCD for Mn$^{2+}$ $d^5$ in a kagome lattice, which has a negligibly small $\langle T_\zeta \rangle$, thereby dismissing it as the origin of the effect.

Also the spin-polarized relativistic (SPR) Korringa-Kohn-Rostoker (KKR) calculations for Mn$_3$Ir and Mn$_3$Ge by Wimmer {\it et al.}~\cite{Wimmer2019} confirmed a non-zero XMCD. It has been suggested that this is connected to a chirality-driven orbital magnetic moment \cite{Wimmer2021}. The scalar spin chirality is closely related to the emergent magnetic
field, a natural concept in the geometric theories of the Hall effect and orbital magnetization \cite{dosSantos2016}. It however requires a non-coplanar rather than a non-colinear magnetic structure   \cite{Smejkal2018}.
In non-coplanar structures, the anomalous Hall effect (AHE) has been associated through the Berry phase with spin chirality \cite{Ye1999, Tatara2002, Taillefumier2006, Chen2014}.
For diluted magnetic systems it was shown that if three spins $S_0$, $S_1$, and $S_2$ are non-coplanar, they contribute to the AHE with a term which is proportional to the scalar chirality $S_0 \cdot (S_1 \times S_2)$ \cite{Tatara2002}. 
This mechanism for AHE  does not involve spin-orbit interaction, but requires only the existence of non-coplanar (chiral) spin configurations.

                           %%%.   Begin FIG. 1. Kagome   %%%%%
\begin{figure}
 	\includegraphics[width = 0.4\textwidth]{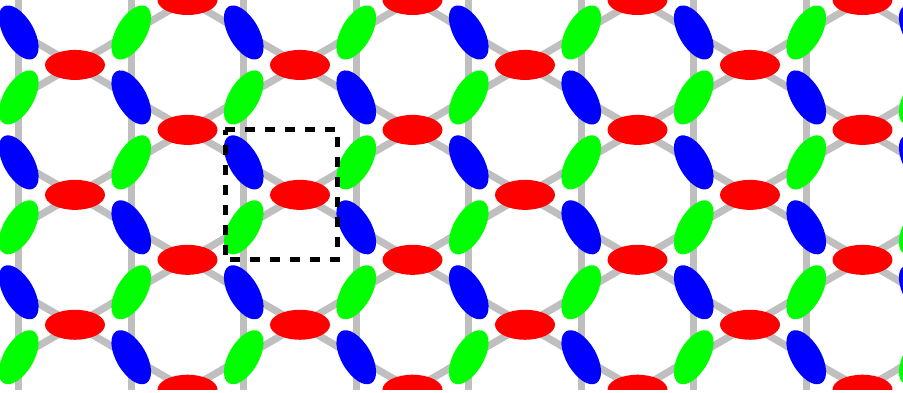}
	\caption{Structure of 2-dimensional kagome lattice with  3-fold symmetry axes and  corner-sharing triangles.
	Only the magnetic atoms are shown.
	%The lattice structure has $D_{2h}$ point group symmetry. 
	In the spin-oriented state the red, blue, and yellow atoms have different spin directions. 
	The dashed box indicates a unit cell of the magnetic structure. 	}
	\label{fig:kagome}
 \end{figure}         %%%.   End FIG. 1.  lKagome   %%%%%

The structure of a 2-dimensional kagome lattice with 3-fold symmetry axes is displayed in Fig.~\ref{fig:kagome}.
The dashed box marks a unit cell of the magnetic structure.
In the spin-oriented state the red, blue, and green atoms have different spin directions. 
With the spin of each atom rotated by $120^\circ$, the AFM structure has no net spin moment. 
Naively, one might then expect a zero XMCD, as is indeed so if the structure has positive chirality (defined as  spin and atomic orientation  having the same sense of rotation). However, this is not necessarily true for negative helicity (defined as  spin and atomic orientation having opposite  sense of rotation). 
As a relevant example, Mn$_3$S  is a  hexagonal AFM with space group $P6_3$/$mmc$ \cite{Brown1990}.
Below the N\'{e}el temperature $T_{\mathrm{N}}$ $\approx$  430 K, the combination of intersite AFM and Dzyaloshinskii-Moriya interactions leads to a 120$^\circ$ spin structure with a uniform negative spin chirality of the in-plane Mn moments because of geometrical frustration. It exhibits a large anomalous Hall conductivity \cite{ Nakatsuji2015} and MOKE \cite{Higo2018} despite the absence of a net magnetic moment.

From the viewpoint of XMCD several questions remain to be addressed:
What is the origin of the non-zero XMCD in the 3-fold symmetric AFM structure?
Is this effect also present for other $n$-fold AFM structures, notably skyrmions?
Can the XMCD be used to deduce the spin direction?
What roles are played by the orbital moment and the magnetic dipole term of the atoms?
 These questions will be answered in the following.

\section{Occurrence of XMCD \label{sec:theory}}

\subsection{Angular dependence}

In x-ray absorption spectroscopy (XAS) at the $L_{2.3}$ edge of a $3d$ transition metal a $2p$ core electron is excited into an unoccupied $3d$ state. 
The participation of the core hole makes that the excitation process  is localized, i.e., restricted to the excited atom in the ligand field of the neighboring atoms. 
The XMCD is obtained as the difference between two XAS spectra  with opposite light helicities.

A general description of the angular dependence of the XMCD in three dimensions  can be found elsewhere~\cite{vanderLaan1998,vanderLaan2010},
but {\color{black} as the minimal model to study the coplanar case two dimensions are sufficient.}
For an isotropic system the XMCD signal is proportional to $\hat{\bf{P}}  \cdot \hat{\bf{S}}$, where $\hat{\bf{P}}$ is the light helicity vector, which is always along the beam direction, and $\hat{\bf{S}}$ is the spin vector.  
In  a lower than octahedral environment of the atom the XMCD will be anisotropic with respect to the atomic axes.
With the spin $\hat{\bf{S}}$ in the $x$-$y$ plane the XMCD spectrum can then be written as 
\begin{equation}
\label{eq:vectors}
I_{\mathrm{XMCD}}    
=   I_x (\hat{\bf{P}} \cdot \hat{\bf{x}}) (\hat{\bf{x}} \cdot \hat{\bf{S}}) 
      + I_y (\hat{\bf{P}} \cdot \hat{\bf{y}}) (\hat{\bf{y}} \cdot \hat{\bf{S}})  ,
\end{equation}
\noindent
where $I_x$ and $I_y$ are the energy-dependent XMCD signals for $\hat{\bf{P}} \parallel \hat{\bf{S}}$ along the local $\hat{\bf{x}}$ and $\hat{\bf{y}}$ directions of the excited atom, respectively. 
{\color{black} The intensity $I \equiv I(E)$ is at an arbitrary photon energy.}

{\color{black} 
The intensity of each peak in the XMCD multiplet can have a different angular dependence. Equation (\ref{eq:vectors}) accounts for this by using a linear combination of so-called fundamental spectra. For electric-dipole transitions in a 2D system there can be only two fundamental spectra,  which we take as  $I_x(E)$ and $I_y(E)$ (for electric-quadrupole transitions there would be more than two spectra). Specifically, the ratio  $I_x/I_y$ will change as a function of photon energy $E$.  An example of the angular dependence of the XMCD  in uniaxial symmetry  described by two fundamental spectra can be found in Ref.~\cite{vanderLaan2010}.}

\begin{figure}
 	\includegraphics[width = 0.3\textwidth]{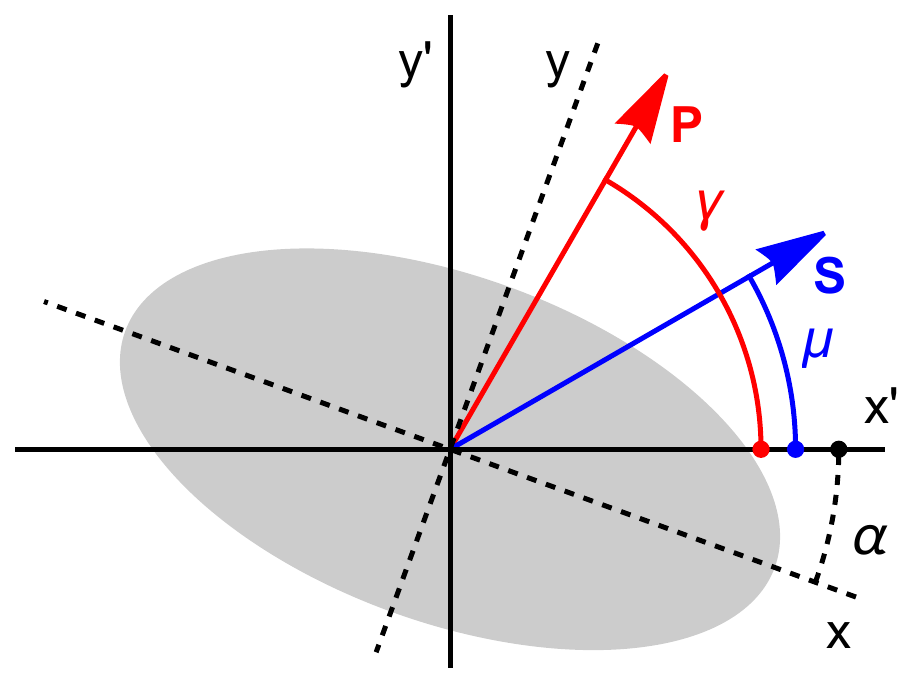}
	\caption{Definition of the angles in the reference frame. The local atom (gray disk) is oriented along $x$, which makes an angle $\alpha$ with the $x'$ axis of the lab frame. The spin direction $\hat{\bf{S}}$ and the helicity vector $\hat{\bf{P}}$  of the circularly polarized x rays are at an angle $\mu$ and $\gamma$, respectively, with respect to $x'$. }
	\label{fig:frame}
 \end{figure}

 With the beam direction fixed, each individual atom $k$ in the kagome unit cell has a different orbital orientation and spin direction. It is assumed that in the local frame  each atom has the same absorption coefficient.
The atom $k$ is oriented along the local $x$ axis, which makes an angle $\alpha_k$ with the lab axis $x'$, so that
Eq.~(\ref{eq:vectors}) can be recast as
\begin{align}
\label{eq:angdep}
I_{\mathrm{XMCD}} ^k  & =    I_x  \cos (\gamma - \alpha_k)  \cos (\mu_k - \alpha_k) \nonumber \\
& + I_y  \sin (\gamma - \alpha_k)  \sin (\mu_k - \alpha_k) ,
\end{align}
where  $\gamma$ and $\mu_k$ are the angles of $\hat{\bf{P}}$ and $\hat{\bf{S}}$, respectively, with respect to $x'$ (see Fig.~\ref{fig:frame}).
Indeed, Eq.~(\ref{eq:angdep}) obeys the correct symmetry properties 
\begin{align}
I_{\mathrm{XMCD}} ^k (\gamma +\pi) & =  - I_{\mathrm{XMCD}} ^k  (\gamma) , \nonumber \\
I_{\mathrm{XMCD}} ^k (\mu_k +\pi) & = - I_{\mathrm{XMCD}} ^k (\mu_k) , \nonumber \\
I_{\mathrm{XMCD}} ^k (\alpha_k +\pi) & = + I_{\mathrm{XMCD}} ^k (\alpha_k) , \nonumber 
\end{align}
for reversal  of the light direction, spin direction, and atomic orientation, respectively.
 
                           %%%.   Begin FIG. 2.   Frame.   %%%%%
          %%%.   End FIG. 2.   Frame.   %%%%%

                           %%%.   Begin FIG. 3. Cells  %%%%%
\begin{figure}
 	\includegraphics[width = 0.5\textwidth]{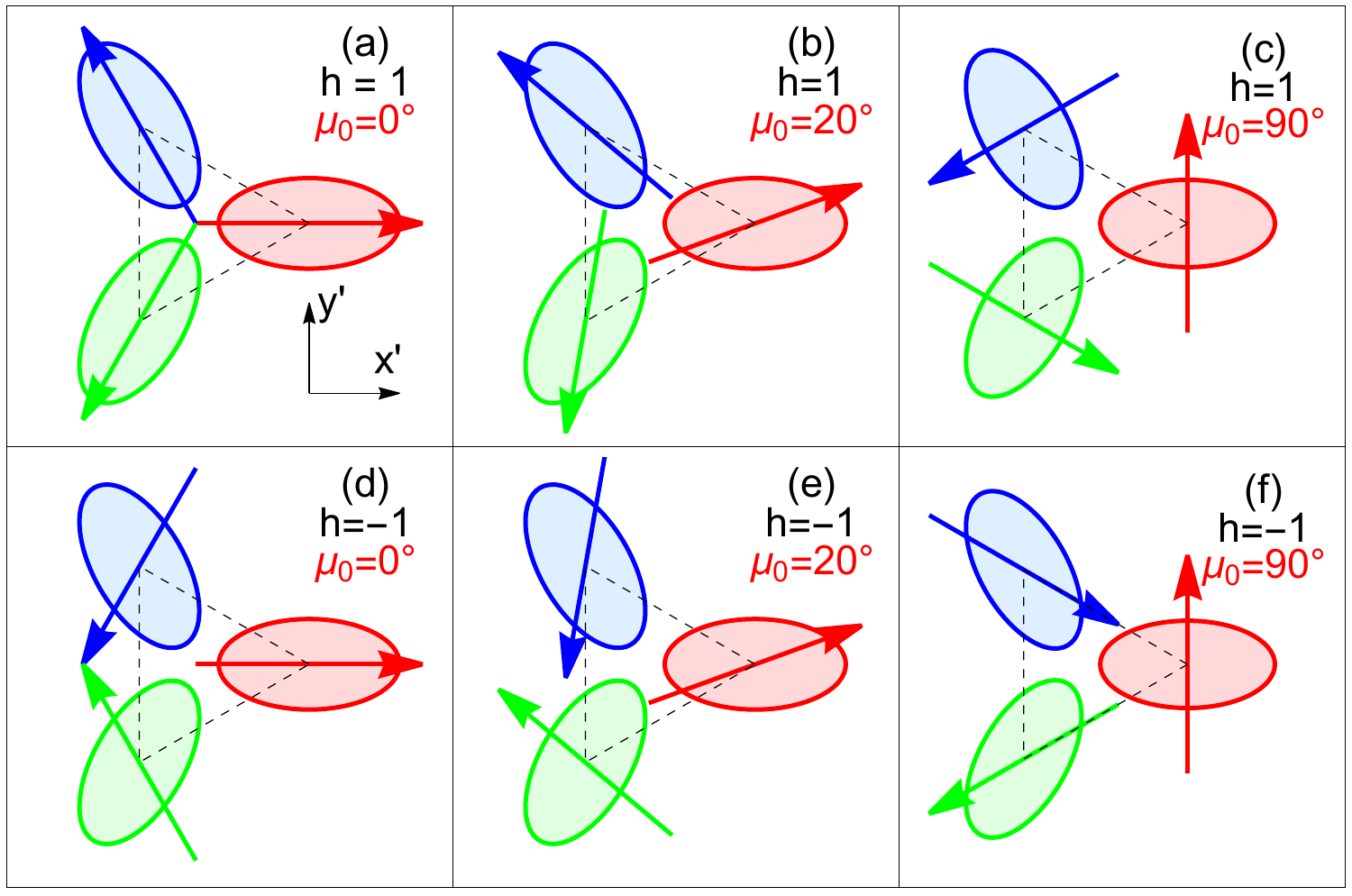}
	\caption{ Magnetic  unit cells  oriented along $\alpha_0$ = $0^\circ$ with positive chirality ($h=1$) for
	(a) $\mu_0 = 0^\circ$ ($A_2$ phase), (b) $\mu_0=20^\circ$, and (c) $\mu_0=90^\circ$  ($A_1$ phase), 
	and with negative chirality ($h=-1)$ for
	(d) $\mu_0 = 0^\circ$, (e) $\mu_0=20^\circ$, and (f) $\mu_0=90^\circ$. \\
		}
	\label{fig:cells}
 \end{figure}    %%%.   End FIG. 3. Cells %%%%%

In the absence of anisotropy $(I_x = I_y)$, Eq.~(\ref{eq:angdep}) reduces to
$I_{\mathrm{XMCD}} ^k = I_x  \cos (\gamma - \mu_k)$, which summed over $k$ for an AFM lattice yields a net zero XMCD.
Thus, it can already be stated that some form of anisotropy would be a necessary condition to have a non-zero total XMCD.

The magnetic unit cell of the kagome lattice, shown in Fig.~\ref{fig:kagome}, can have different
spin chiralities  as depicted in Fig.~\ref{fig:cells}. 
Starting from the $x'$ axis, the three atoms and their spins, with angles  $\alpha_k$ and $\mu_k$, are labeled $k$ = 0, 1, 2 in the counterclockwise convention, and colored red, blue, and green, respectively.
Taking  $\alpha_0 = 0^\circ$, examples are shown with  $\mu_0 = 0^\circ$, $ 20^\circ$, and $90^\circ$.
In the case that $\alpha_0 \ne 0^\circ$ we are free to rotate the unit cell to  $\alpha_0 = 0^\circ$, i.e., align along the $x'$ axis.
For positive chirality ($h = 1$)  the atomic orientation and spin direction in Fig.~\ref{fig:cells} follow the same sense of rotation, increasing by 120$^\circ$ between the respective atoms in the triangle.
In contrast, for  negative chirality  ($h = -1$) the spin direction has the opposite sense of rotation as the  atomic orientation.

For positive chirality, the 3-fold rotational symmetry of the structure leads to a vanishing total XMCD.
For, e.g., Figs.~\ref{fig:cells}(a) and \ref{fig:cells}(b), Eq.~(\ref{eq:angdep}) gives for the atom  $k$, 
\begin{align}
& I_{\mathrm{XMCD}}^k =   I_x    \cos ( \gamma - \alpha_k ) ~~~ \mathrm{for } ~ \mu_k = \alpha_k ,   \nonumber \\  
& I_{\mathrm{XMCD}}^k =   I_y    \sin ( \gamma - \alpha_k ) ~~~ \mathrm{for } ~ \mu_k = \alpha_k + 90^\circ ,   
\end{align}
respectively. Since $\gamma$ is fixed with respect to $x'$ whereas $\alpha_k$ is summed over $k$, the total XMCD amounts to zero. 
However, for negative chirality  the combination of atomic orientation and spin direction lacks rotational symmetry. For completeness the symmetry properties of the spin vectors are given in Appendix \ref{app:A}, although no explicit use will be made of group-theoretical arguments.  Instead a more transparent analytical derivation is given.

\subsection{General derivation}
\label{sec:derivation}

Equation~(\ref{eq:angdep}) can be recast into a sum over an isotropic and an anisotropic part as
\begin{align}
\label{eq:angdep2}
 I_{\mathrm{XMCD}} ^k & = \frac{1}{2} (I_x + I_y)  \cos (\gamma - \mu_k)    \nonumber   \\
  & +  \frac{1}{2} (I_x - I_y)   \cos (  \gamma + \mu_k - 2 \alpha_k) .
\end{align}
Both chiralities can be captured by taking the angles for atom $k$ as
\begin{align}
\label{eq:h}
\alpha_k & = \alpha_0 + \frac{2 \pi}{3}  k ,  \nonumber \\
\mu_k & = \mu_0 +  \frac{2 \pi}{3}  k  h ,
\end{align}
where $h$ = +1 and $-1$ for positive and negative chirality, respectively. This gives
\begin{align}
 I_{\mathrm{XMCD}} ^k & = \frac{1}{2} (I_x + I_y)  \cos (\gamma - \mu_0 -  \frac{2 \pi}{3}  k  h)  \nonumber \\
   + &  \frac{1}{2} (I_x - I_y) \cos [  \gamma + \mu_0 - 2 \alpha_0 +  \frac{2 \pi}{3}  k (h - 2) ]   .
\label{eq:key}
\end{align}
\noindent
Summing over $k = 0, 1, 2$ yields our key result, the total XMCD over the unit cell,
\begin{align}
\label{eq:Itotneg}
I_{\mathrm{XMCD}} ^{\mathrm{total}} & = 0 ~~ \mathrm{for} ~ h=+1,    \nonumber \\
I_{\mathrm{XMCD}} ^{\mathrm{total}} & = \frac{3}{2} (I_x - I_y) \cos ( \gamma  + \mu_0 - 2 \alpha_0 )  ~~\mathrm{for} ~ h=-1 , 
\end{align}
where  $\gamma$,  $\mu_0$, and $\alpha_0$ have  fixed values in the lab frame.
Not surprisingly, the isotropic part has dropped out.

It can  be immediately verified from Eq.~(\ref{eq:Itotneg}) that if $\alpha_0$ and  $\mu_0$ are increased by $120^\circ$ and $-120^\circ$, respectively, then the intensity remains  unchanged. Therefore, the anisotropic part for each of the three atoms is the same, so that the total XMCD is 3$\times$ the individual atom contribution.

Equation~(\ref{eq:Itotneg}) shows that  for negative chirality the total XMCD does not vanish unless  the XMCD of the local atom is isotropic ($I_x = I_y$) or when $ \gamma = 2 \alpha_0  - \mu_0  \pm 90^\circ$, which corresponds to the zero crossings between positive and negative dichroism.

                           %%%.   Begin FIG. 4.  Angular dependence %%%%%
\begin{figure}
 	\includegraphics[width = 0.48\textwidth]{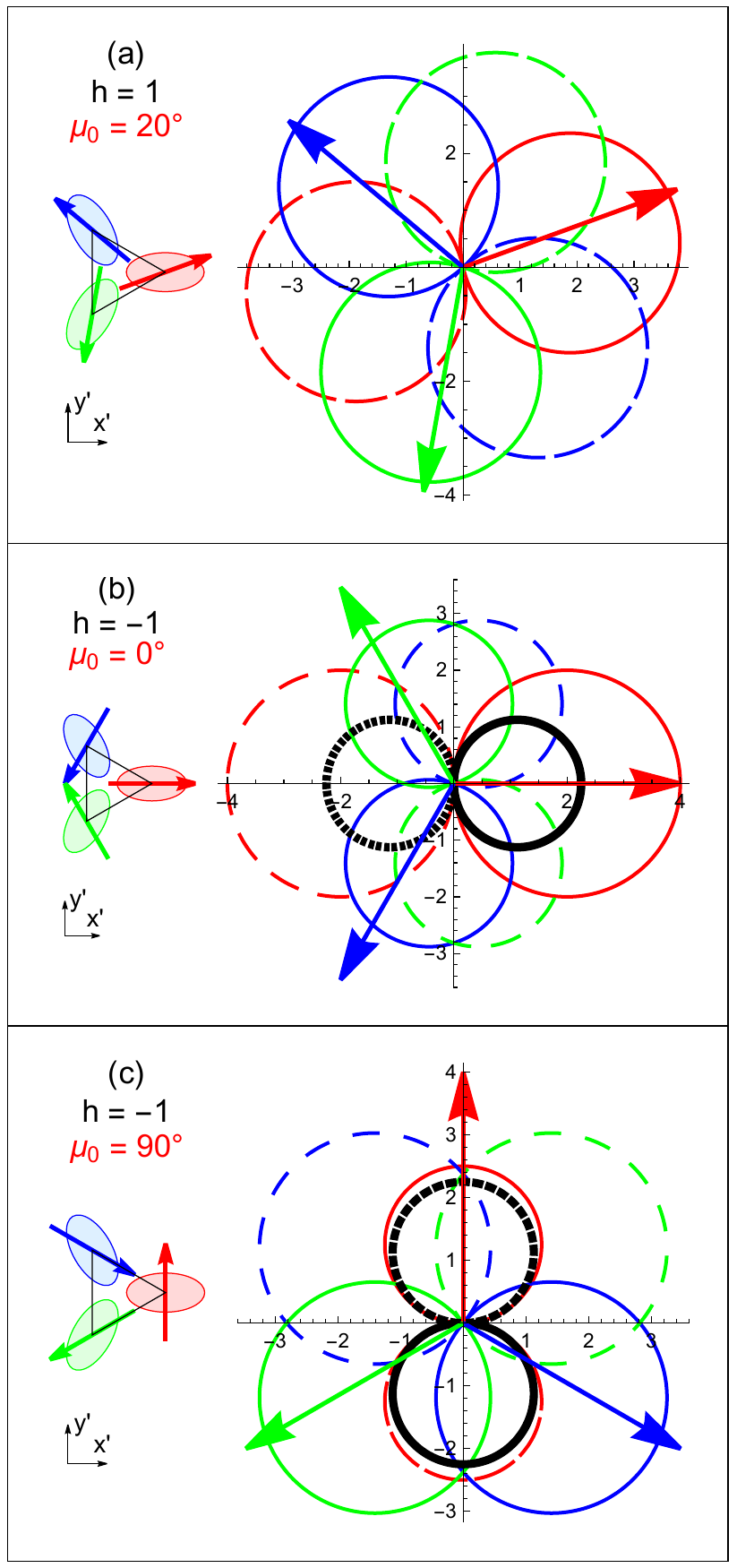}
	\caption{  The magnetic unit cell,  oriented at  $\alpha_0 = 0^\circ$, is shown in the inset at the left.
	The main image shows the angular dependence of the XMCD as a function of $\gamma$. The red, blue, and green curves give the XMCD for the three magnetic atoms with spin directions shown by the color-coded arrows.
	The total XMCD is given by the thick black curve.
The plots have been obtained from Eq.~(\ref{eq:key}) for $I_x=4$, $I_y=2.5$, and the full and dashed curves correspond to positive  and negative signals, respectively. 
	(a) Positive helicity ($h=1$) with $\mu_0 = 20^\circ$, which gives a total XMCD = 0.
	(b) Negative helicity ($h= -1$) with $\mu_0 = 0^\circ$, which gives a total XMCD = $ \frac{3}{2} (I_x-I_y) \cos \gamma$, with maximum at $\gamma = -\mu_0 = 0^\circ$, i.e., along $+x'$.
	(c) Negative helicity with $\mu_0 = 0^\circ$, which gives a total XMCD = $-\frac{3}{2} (I_x-I_y) \sin \gamma$, with maximum  at  $\gamma = -\mu_0 = -90^\circ$,  i.e., along $-y'$. 	
	 	}
	\label{fig:ang-dep}
 \end{figure}    %%%.   End FIG. 4. Angular dependence.  %%%%%

Further insight can be gained from a graphical illustration of the angular dependence  of the XMCD  as a function of the incident beam angle $\gamma$. Choosing for convenience $I_x > I_y$ in Eq.~(\ref{eq:key}),  Fig.~\ref{fig:ang-dep} shows angular plots with the lattice oriented at $\alpha_0 = 0^\circ$.  
The red, blue, and green curves give the XMCD for each of the three magnetic atoms with spin directions shown by the color-coded arrows. 
The corresponding magnetic unit cell is shown in the inset on the left.

 Figure~\ref{fig:ang-dep}(a) shows a unit cell with positive helicity ($h=1$) and $\mu_0 = 20^\circ$. It demonstrates the 3-fold rotational symmetry of the angular dependence, whereby the total XMCD vanishes.
 
Figure~\ref{fig:ang-dep}(b) shows a unit cell with negative helicity and $\mu_0 = 0^\circ$. Here, the magnitude of the XMCD for the red atom is larger than that for the blue and green atoms, which is due to their different value for the angle $| \mu_k - \alpha_k |$. The total XMCD is shown by the thick black curve, and since $\alpha_0 = \mu_0 =0^\circ $ this gives $I_{\mathrm{XMCD}}^{\mathrm{total}} = \frac{3}{2} (I_x-I_y) \cos \gamma$, which has a maximum  at $\gamma = 0^\circ$.

It becomes  interesting when $\mu_0 \ne 0$, in which case the maximum dichroism is no longer oriented along $S_0$ (the red arrow) since $\gamma = - \mu_0$. 
Figure~\ref{fig:ang-dep}(c) shows a unit cell with $\mu_0 = 90^\circ$. The magnitude of the XMCD for the red atom is smaller than that for the blue and green ones. 
The total XMCD is
$I_{\mathrm{XMCD}}^{\mathrm{total}}  = -\frac{3}{2} (I_x-I_y) \sin \gamma$ with the maximum  at $\gamma = - 90^\circ$.

Both examples in Figs.~\ref{fig:ang-dep}(b) and \ref{fig:ang-dep}(c) are consistent  with the cluster calculations for the $L_{2,3}$ XMCD of Fe$^{2+}$ $d^6$ and Mn$^{2+}$ $d^5$ in a kagome lattice, which were carried out by Sasabe {\it et al.}~\cite{Sasabe2021} using a fixed geometry.

                           %%%.   Begin FIG. 5.   ---  20 Degrees %%%%%
\begin{figure}
 	\includegraphics[width = 0.48\textwidth]{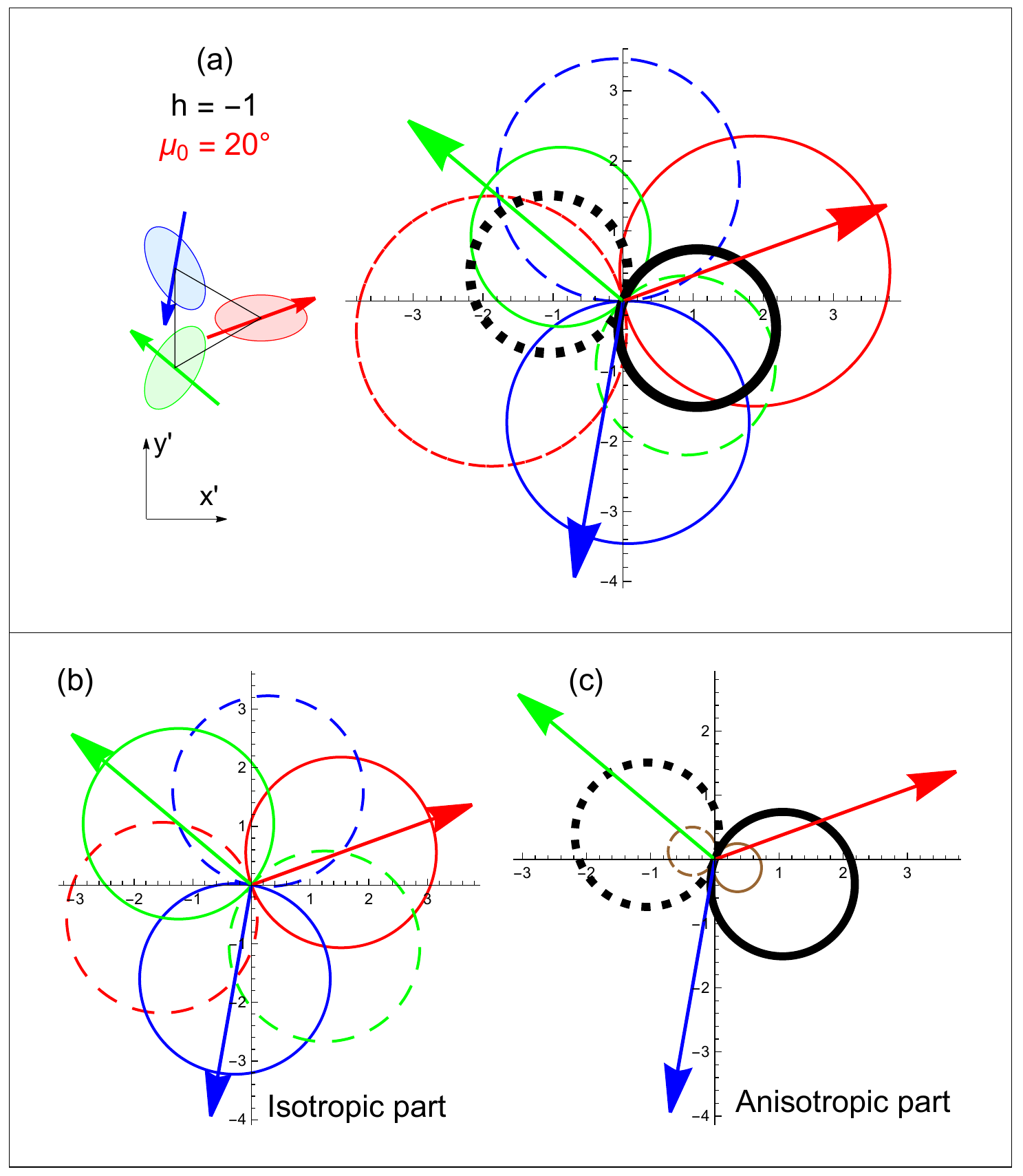}
	\caption{ 
	(a) The angular dependence as a function of $\gamma$ for negative helicity with $\mu_0 = 20^\circ$. The description of the plot  is as in Fig.~\ref{fig:ang-dep}.
	The total XMCD =
	$ \frac{3}{2} (I_x-I_y) \cos (\gamma+20^\circ)$, which has a maximum for $\gamma = - \mu_0 = -20^\circ$, thus 
	at the opposite site of  the $x'$ axis than $S_0$  (red arrow).
	The lower panel shows the decompostion of the XMCD in isotropic and anisotropic part. 
	(b) The isotropic part of the XMCD, which shows no total signal.
	(c) The anisotropic part, where all three magnetic atoms have the same angular dependence, shown by the brown curve and which together add up to the black curve.
	}
	\label{fig:20-deg}
 \end{figure}    %%%.   End FIG.  5.   ---  20 Degrees %%%%%

Finally, Fig.~\ref{fig:20-deg} for  $h=-1$ and $\mu_0=20^\circ$ reveals the intricacies of the anisotropy in the XMCD signal.
For each of the atoms the angular dependence of the XMCD,  shown in Fig.~\ref{fig:20-deg}(a), is separated into its isotropic and anisotropic part.
 The isotropic part in Fig.~\ref{fig:20-deg}(b) shows a 3-fold rotational symmetry and thus has no total XMCD.
 In the anisotropic part shown in Fig.~\ref{fig:20-deg}(c), all three atoms give the same angular dependent intensity, which is shown by the brown curve and together these add up to the black curve.
The total XMCD is	$ \frac{3}{2} (I_x-I_y) \cos (\gamma+20^\circ)$, with a maximum at $\gamma = - \mu_0 = -20^\circ$,  thus 
at the opposite site of  the $x'$-axis than $S_0$. 
 
For a non-aligned lattice orientation ($\alpha_0 \ne 0^\circ$), the maximum of the dichroism is  found at  $\gamma = 2 \alpha_0  - \mu_0$.
In the experiment, one can only determine the spin direction within this constraint, whereby different geometries can result in the same angular dependence.
For instance, rotating the $x'$ axis by $90^\circ$ in Fig.~\ref{fig:ang-dep}(c)  so that $\alpha_0 = 90^\circ$ and $\mu_0 = 180^\circ$ or
 rotating the $x'$ axis by $20^\circ$ in Fig.~\ref{fig:20-deg}(a)   so that $\alpha_0 = 20^\circ$ and $\mu_0 = 40^\circ$ 
 gives in both cases  exactly the same angular dependence as in Fig.~\ref{fig:ang-dep}(b) for $\alpha_0 =  \mu_0 = 0^\circ$.
Hence,  the angular dependence of the beam direction can be used to  obtain  the spin direction only if  the atomic orientation is known.
Note that this also means  that in the experiment particular care should be taken  with the azimuthal alignment of the sample.

\subsection{Unit cell with multifold axis}

Next, we consider whether a unit cell with $n$-fold rotational symmetry, $C_{nv}$ ($n \ge 2$) can have a non-vanishing total XMCD.
Replacing $ \frac{2 \pi}{3} $ by $ \frac{2 \pi}{n}$ in Eq.~(\ref{eq:key}) and summing  $k$ from 0 to $n-1$ gives that  $I_{\mathrm{XMCD}}^{\mathrm{tot}} = 0$  except for the above-treated case of ($n = 3$, $h=-1$).
This also applies to $n=6$, which can be considered as the sum of two lattices with $n=3$ that are rotated 60$^\circ$ with respect to each other. 
According to Eq.~(\ref{eq:Itotneg}) the XMCD of these two unit cells have opposite signs and thus cancel each other out. 

Skyrmions in crystalline lattices of B20-type compounds have 6-fold symmetry  axes and skyrmions in metallic multilayers might be regarded as having ($n \to \infty$) fold symmetry, so that none of these will qualify for any total XMCD. However, resonant elastic x-ray scattering (REXS) has shown to be able to unambiguously resolve the chirality, thanks to the interference effects between the different sites \cite{Zhang2017}. 

\section{Spin and orbital ground state moments}

\subsection{Orbital moment}

According to the orbital magnetic sum rule the expectation value of the ground state orbital moment $\langle L_\zeta \rangle$  is proportional to the energy-integrated XMCD signal  along the $\zeta$ direction \cite{Thole1992,vanderlaan1998sr}.
For the atom $k$ the angular dependence  is 
\begin{align}
\label{eq:angdepL}
\langle L ^{k } \rangle  =  & \frac{1}{2} \left[\langle L_x \rangle + \langle L_y \rangle \right]  \cos (\gamma - \mu_k)  \nonumber \\
 + & \frac{1}{2} \left[ \langle L_x \rangle - \langle L_y \rangle  \right]    \cos (\gamma  +  \mu_k - 2 \alpha_k) ,
\end{align}
\noindent
which agrees with Eq.~(35) of Ref.~\cite{vanderLaan1998}. Since this has the same angular dependence as the XMCD signal in Eq.~(\ref{eq:angdep2}), it can  only be non-zero for a 3-fold lattice  with negative chirality.
Summation over $k$ gives the total angular dependence for $h=-1$ as
\begin{equation}
\langle L _{\mathrm{total}} \rangle 
= \sum_k\langle L^k \rangle 
= \frac{3}{2} [\langle L_x \rangle - \langle L_y \rangle ]  \cos (\gamma  +  \mu_k - 2 \alpha_k) . 
\label{eq:Lzneg}
\end{equation}
Importantly, the obtained quantity is given by the anisotropy in the atomic orbital moment and not by the orbital moment itself.

\subsection{Magnetic dipole term}

A second sum rule relates the effective spin moment $\langle S_{\mathrm{eff},\zeta}  \rangle$ along the $\zeta$-direction  of a $3d$ metal to the energy-integrated XMCD intensity over the $L_3$ absorption
edge minus twice that over the $L_2$ edge \cite{Carra1993,vanderlaan1998sr}.
$\langle S_{\mathrm{eff},\zeta}  \rangle =  \langle S_\zeta \rangle + \frac{7}{2} \langle T_\zeta \rangle$, where $\langle T_\zeta \rangle$ is the expectation value of the magnetic dipole term $ \mathbf{T} = \mathbf{\hat{S}} -3  \mathbf{\hat{r}} ( \mathbf{\hat{r}} \cdot \mathbf{S})$, with  $ \mathbf{\hat{r}}$  the position unit vector \cite{Carra1993}.
While the spin moment $\mathbf{S}$ is isotropic, $\mathbf{T}$ gives the anisotropy of the spin moment due to the coupling with the charge quadrupole moment and spin-flip terms.
In agreement with Eq.~(39) of Ref.~\cite{vanderLaan1998}, $\langle T \rangle$ has 
  the same angular dependence as the anisotropic orbital moment, so that  for $h=-1$,
\begin{equation}
\langle S_{\mathrm{eff}}  \rangle
= \sum_k \langle S_{\mathrm{eff}}^k  \rangle
= \frac{21}{4} [\langle T_x \rangle - \langle T_y \rangle ]  \cos (\gamma  +  \mu_k - 2 \alpha_k) , 
\label{eq:Lzneg}
\end{equation}
where the isotropic spin moment has dropped out. 
If {\bf{T}} is taken as a quadrupole moment along the $x$-axis then $\langle T_y \rangle = - \frac{1}{2} \langle T_x \rangle$.

It should be stretched that,  contrary to earlier proposition  \cite{Yamasaki2020},  the presence of {\bf{T}} is not the unique origin of a non-zero total XMCD for negative chirality structures.
{\color{black} 
The sum rule measures $\langle T \rangle$ in the initial state and this can be zero, such as for the Hund's rule ground state of Mn $3d^5$. The anisotropy in the XMCD is then caused by the extra $d$ electron in the final state.
For instance, anisotropic spectra for an isotropic ground state have been reported for Ti $3d^0 \to 2p^53d^1$ \cite{Arenholz2010}. }

\section{conclusions}
It is shown most generally using a straightforward analytical derivation as well as by graphical illustration that a triangular  structure with negative spin chirality allows the existence  of a non-zero total XMCD, despite the fact that  the total spin moment vanishes.
A necessary condition is an anisotropic XMCD  signal $(I_x \ne I_y)$ for the local atom.

The non-zero XMCD for negative spin chirality can be understood in a simple way:
The anisotropic part of the XMCD signal depends on the angle
$ (2 \alpha_k - \mu_k)$ with respect to the incident beam direction. %  [Eq.~(\ref{eq:Itotneg})] 
This angle is invariant ($\pm 360^\circ$) 
for a 3-fold rotation in which $\alpha_k$ and $\mu_k$ are rotated by $120^\circ$ and  $-120^\circ$, respectively  (see Fig.~\ref{fig:frame} for angle description). 
This only holds  for 3-fold symmetric lattices, and  a zero value is returned for any other $n$-fold symmetry.

The extrema of the  total dichroism are found at $\gamma = 2 \alpha_0  - \mu_0 $ and  $2 \alpha_0  - \mu_0 + 180 ^\circ$.
Thus,  for $\alpha_0 \ne 0$ the maximum dichroism is not aligned along the spin direction, but also depends on the relative orientation with respect to the atomic orientation. This also implies that if in the experimental setup we do not know $\alpha_0$ then a particular spin direction has no unique angular dependence. 

By the x-ray magneto-optical sum rules, the orbital and spin moments  are proportional to the integrated intensities of the XMCD\@.  Because the total XMCD of the 3-fold lattice is related to the anisotropic XMCD of the single atom, the integrated intensities yield the anisotropic part of the orbital and effective spin moment, which correspond to $ \langle L_x \rangle - \langle L_y \rangle $ and  magnetic dipole term, respectively.
Both display the same angular dependence as the total XMCD.
Note that even when the energy-integrated XMCD is zero, the XMCD signal can still be positive and negative in equal amounts accross the photon energy range of the absorption edge.
Therefore, the presence of an anisotropic orbital moment or a magnetic dipole term is not conditional  for a non zero XMCD signal.
For instance, in the x-ray transition  $3d^5 \to 2p^5 3d^6$ 
the final state can induce an anisotropy in the XMCD spectrum, while the  orbital moment and magnetic dipole term in the Hund's rule ground state  $3d^5$ are zero.

While the presented results confirm earlier cluster calculations \cite{Sasabe2021} and SPR-KKR calculations \cite{Wimmer2019} in fixed geometries, here we  developed a general rule under which this effect occurs without the need for any detailed spectral calculations. 
Importantly, we showed that the effect exists in a coplanar geometry and that  non-coplanar spin  moments are not required for its existence. 

{\color{black}
We have not discussed the situation in three dimensions, in which case Eq.~(\ref{eq:vectors}) also contains a term 
 $I_z (\hat{\bf{P}} \cdot \hat{\bf{z}}) (\hat{\bf{z}} \cdot \hat{\bf{S}}) $.
 If the spin $ {\bf{S}}$ of each atom is in-plane, then this term is zero, and for the other terms we can simply take the in-plane components of $ {\bf{P}}$. If $  {\bf{S}}$ also has an out-of-plane component, then there are three fundamental spectra, which makes an analytical analysis less transparent, and beyond the scope of the paper. A numerical calculation would then be more appropriate. Experimentally, one might be able to perform measurements with $ \hat{\bf{P}}$ at different angles of incidence, especially  $ \bf{P} \! \parallel \! \bf{z}$, to determine the XMCD contribution from the spin component along the $z$ direction. }

Our findings are useful because antiferromagnets have unique attributes that are not found in ferromagnets, such as high-frequency dynamics, robustness against external perturbations, and absence of stray field, giving rise to new patterns of  antiferromagnetic spintronics \cite{ Baltz2018,Siddiqui2020} and topological antiferromagnetic spintronics \cite{Smejkal2018}.

\appendix

\section{Spin configurations on triangle}
 \label{app:A}

The AFM spin structures in Fig.~\ref{fig:cells} are subjected to the point group symmetry $C_{3v}$, and their irreps and basis vectors are given in Table \ref{tab:irreps}. The labeling follows the notation in Refs.~\cite{Essafi2017,Benton2021}. The character table for the group $C_{3v}$ is given in Table \ref{tab:C3v}. Note that the spin is an axial vector. Kramers theorem tells that by time reversal each triangular structure  can also have all its spins reversed.
For negative helicity the two irreps $E$ share the same symmetry properties and can be mixed.

\begin{table}[h]                %%%% TABLE 1
\begin{ruledtabular}
\begin{tabular}{lcccccc} 
Fig. & $h$ 	&  $\mu_0$ & Irreps & $S_0$	&  $S_1 $  &	$S_2$ 	\\
\hline
 \ref{fig:cells}(a) & $+1$	& $0^\circ$ & $S_\perp (A_2)$ &   $(1,0)$	& $ ( -\frac{1}{2}, \frac{1}{2} \sqrt{3} ) $  &	$ (-\frac{1}{2} ,  -\frac{1}{2} \sqrt{3} ) $ 	\\
 \ref{fig:cells}(c) & $+1$	&   $90^\circ$ &$S(A_1)$ &   $(0,1)$	&  $ ( -\frac{1}{2} \sqrt{3} ,  -\frac{1}{2} ) $   &	$ ( \frac{1}{2} \sqrt{3} , -\frac{1}{2}) $  	\\
 \ref{fig:cells}(d)	& $-1$ & $0^\circ$ & $S_{2,\mathrm{AF}}(E)$ &   $(1,0)$	& $ ( -\frac{1}{2},  -\frac{1}{2} \sqrt{3} ) $  &	$ (  -\frac{1}{2} , \frac{1}{2} \sqrt{3}) $  	\\
 \ref{fig:cells}(f)	& $-1$ &  $90^\circ$ &  $S_{1,\mathrm{AF}}(E)$ &  $(0,1)$	& $ ( \frac{1}{2} \sqrt{3} , -\frac{1}{2}) $   &	$ (  -\frac{1}{2} \sqrt{3} , -\frac{1}{2} ) $ 	\\
\end{tabular}
\caption{ Irreps for the AFM spin structures shown in Fig.~\ref{fig:cells}  with the three basis vectors $S_k = (\cos \mu_k, \sin \mu_k)$, which are the spins in counterclockwise order located on the red, blue, and green
 atoms, respectively.}
\label{tab:irreps}
\end{ruledtabular}
\end{table}

\begin{table}[h]                %%%% TABLE 2
\begin{ruledtabular}
\begin{tabular}{lrrr} 
$C_{3v}$ &  $e$	&  $C_3$	&  $\sigma_v$ 		\\
\hline
$A_1$ &  1	&  1	& 1 	\\
$A_2$ &  1	&  1	& $-1 $	\\
$E$     &  2	&  $-1$	& 0 	\\
\end{tabular}
\caption{ Character table for the point group  symmetry $C_{3v}$, with the conjugacy classes  under the $e$ identity operator, $C_3$ rotations, and $\sigma_v$ reflections.  The irreps $A_1$, $A_2$, and $E$ are in Mulliken notation.}
\label{tab:C3v}
\end{ruledtabular}
\end{table}

% \bibliography{references}

%apsrev4-2.bst 2019-01-14 (MD) hand-edited version of apsrev4-1.bst
%Control: key (0)
%Control: author (8) initials jnrlst
%Control: editor formatted (1) identically to author
%Control: production of article title (0) allowed
%Control: page (0) single
%Control: year (1) truncated
%Control: production of eprint (0) enabled
%

\end{document}